\documentclass[aps,prd]{revtex4-1}
\pdfoutput=1
\usepackage{subfigure,graphics} 
\usepackage{flafter} 
\begin{document} 
\title{
Cosmological bounds on tachyonic neutrinos}
\author{P. C. W. Davies}
\email{paul.davies@asu.edu}
\affiliation{Beyond Center for Fundamental Concepts in Science, 
Arizona State University, P.O. Box 871504, Tempe, AZ  85287, USA}
\author{Ian G. Moss}
\email{ian.moss@ncl.ac.uk}
\affiliation{School of Mathematics and Statistics, Newcastle University, 
Newcastle upon Tyne, NE1 7RU, UK}

\date{\today}


\begin{abstract}
Recent time-of-flight measurements on muon neutrinos in the OPERA neutrino oscillation
experiment have found anomalously short times compared to the light travel-times,
corresponding to a superluminal velocity, $v-1=2.37\pm0.32\times 10^{-5}$
in units where $c=1$. We show that cosmological bounds rule out an explanation
involving a Lorentz invariant tachyonic neutrino.  At the OPERA energy scale, 
nucleosynthesis constraints imply $v-1<0.86\times 10^{-12}$ and the Cosmic 
Microwave Background observations imply $v-1<7.1\times 10^{-23}$.
The CMB limit on the velocity of a tachyon with an energy of 10 MeV is 
stronger than the SN1987A limit.
Superluminal neutrinos that could be observed at particle accelerator energy 
scales would have to be associated with Lorentz symmetry violation.
\end{abstract}
\pacs{PACS number(s): }

\maketitle
\section{Introduction}

Recent time-of-flight measurements on muon neutrinos in the OPERA neutrino oscillation
experiment have found anomalously short flight-times compared to the light-travel times
\cite{Adam:2011zb}.
The anomaly corresponds to a superluminal velocity, $v-1=2.37\pm0.32\times 10^{-5}$
in units where $c=1$. The simplest type of particle with superluminal velocities would
be a tachyon, a Lorentz invariant particle with imaginary mass, although this possibility
conflicts with the observations of (anti)neutrinos associated with the supernova 
SN 1987A, which imply $v-1<2\times 10^{-9}$ for 10 MeV neutrinos 
\cite{Hirata:1987hu,Bionta:1987qt,Stodolsky:1987vd,Longo:1987gc,Cacciapaglia:2011ax}.
We show that cosmological bounds support the supernova result and also rule out
a tachyonic explanation of the neutrino timing anomaly. We find, at the OPERA
energy scale, $v-1<0.86\times 10^{-12}$ from nucleosynthesis constraints and
$v-1<7.1\times 10^{-23}$ from the Cosmic Microwave Background (CMB) observations.
The CMB limit on the velocity of a tachyon with an energy of 10 MeV is 
stronger than the SN1987A limit.

\section{Tachyons in an expanding universe}

Our treatment of tachyonic particles in an expanding universe follows Refs.
\cite{Davies:1975} and \cite{Davies:2004eq}. The tachyon is a particle with
imaginary mass $i\mu$ and speed $v>1$, whose energy and momentum in a local 
inertial frame are given by
\begin{eqnarray}
E&=&\mu (v^2-1)^{-1/2},\\
{\bf p}&=&\mu{\bf v}(v^2-1)^{-1/2}.
\end{eqnarray}
We also find it useful to express the speed as a function of the energy,
\begin{equation}
v(E)=\left(1+{\mu^2\over E^2}\right)^{1/2}.\label{speed}
\end{equation}
The tachyon speeds up if its energy is reduced.

Consider such a particle in a spatially-flat Friedmann-Robertson-Walker universe with
cosmological time $t$ and scale factor $a(t)$. If we use $v^i$ to denote the velocity
is the co-moving frame, then the conserved linear momentum is
\begin{equation}
p_i=\mu a^{-2} v^i.
\end{equation}
The energy and momentum of the tachyon are related by
\begin{equation}
p_ip^i-E^2=\mu^2.
\end{equation}
The energy of a tachyon in an expanding universe is therefore given by
\begin{equation}
E=\mu\left({p_ip_i\over\mu^2}{1\over a^2}-1\right)^{1/2}.\label{energy}
\end{equation}
The same result was obtained was obtained in Ref. \cite{Davies:2004eq} using the law
of addition of velocities in an expanding universe. 
The important feature here is that the energy of a non-interacting
tachyon falls {\it faster} than the usual $1/a$ rate for massless particles.
At some point the energy falls to zero and the tachyons disappear from the universe.
This may be interpreted as a tachyon-antitachyon annihilation event \cite{Davies:1975}.

The foregoing analysis only applies to free tachyons. A tachyon in thermal equilibrium with
ordinary particles at temperature $T$ would have energy $E\approx T$, as usual. If the
tachyons come out of equilibrium, then the constants in Eq. \ref{energy} are fixed by
the energy at the relevant decoupling time.

\section{Cosmological bounds}

We shall start off the cosmological bounds with nucleosynthesis constraints 
(see e.g. \cite{Fields:2006ga} and \cite{Steigman:2005uz} for a review),
and focus on events between neutrino decoupling at a temperature $T_{\rm weak}$ and element
formation at $T_{\rm nuc}$. An effective neutrino number $N_\nu$ is defined in terms of the 
energy densities of neutrinos $\rho_\nu$ and photons $\rho_\gamma$. 
For times preceding electron-positron annihilation,
\begin{equation}
{\rho_\nu\over\rho_\gamma}={7\over 8}N_\nu.\label{defn}
\end{equation}
There is an extra factor of $(4/11)^{4/3}$ on the right-hand-side after electron-positron annihilation.
In the standard model, with three lepton families, we have $N_\nu=3$.

The cosmological expansion rate is sensitive to the number of neutrino species, and therefore both
the neutrino decoupling time and the neutron/proton ratio at the time of
helium formation are dependent on $N_\nu$. By modelling the primordial element abundances
and comparing them to observations it is possible to set limits on $\Delta N_\nu=N_\nu-3$.
Typical examples of these limits are $\Delta N_\mu>-1.7$ ($95\%$ CL) \cite{Steigman:2005uz} and
 $\Delta N_\mu>-1.0$ ($95\%$ CL) \cite{Fields:2006ga}. Note that these results assume
$\Delta N_\mu$ is constant from decoupling to helium formation.

Free tachyonic neutrinos, the energy is given by Eq. (\ref{energy}) with $T\propto 1/a$, and we have 
$E_\nu=T_{\rm weak}$ at neutrino decoupling to fix the constants, 
\begin{equation}
E_\nu(T)=\mu\left({T^2\over T_{\rm weak}^2}+{T^2\over \mu^2}-1\right)^{1/2}.
\end{equation}
Neutrino oscillation experiments imply that all three neutrino species would have a 
similar tachyonic mass \cite{GonzalezGarcia:2007ib}. The neutrino energy density $\rho_\nu\propto T^3E_\nu$
and the photon energy density $\rho_\gamma\propto T^4$, so that the effective number of neutrino species 
defined by Eq. (\ref{defn}) is
\begin{equation}
N_\nu(T)=3\left(1+{\mu^2\over T_{\rm weak}^2}-{\mu^2\over T^2}\right)^{1/2}.\label{nnu}
\end{equation}
Note that $N_\nu=3$ at neutrino decoupling, then it decreases to zero and remains zero after 
the tachyon annihilation.

Given limits on $\Delta N_\nu$ at the nucleosynthesis time, then
we can invert (\ref{nnu}) to set limits on the tachyon mass
\begin{equation}
\mu^2\le{2\over 3}{T^2_{\rm nuc}T^2_{\rm weak}\over T^2_{\rm weak}-T^2_{\rm nuc}}
|\Delta N_\nu|.
\label{nul}
\end{equation}
Typical values would be $T_{\rm weak}=0.8{\rm MeV}$ for decoupling, $T_{\rm nuc}=0.1{\rm MeV}$
for light element formation and $|\Delta N_\nu|<2$,
\begin{equation}
\mu\leq 0.12{\rm MeV}.
\end{equation}
The bounds on the tachyonic mass translate into bounds on the speed of the tachyon at a given energy scale
via Eq. (\ref{speed}). If we where to compare with the OPERA results \cite{Adam:2011zb}, at an energy of  
$28{\rm GeV}$,
\begin{equation}
v(28{\rm GeV})-1<0.86\times 10^{-12}.
\end{equation}
The speed reported by the OPERA announcement corresponds to $v-1=2.4\times 10^{-5}$.

CMB observations can be used to give information about the neutrino density 
of the universe at later times than the nucleosynthesis era. The acoustic oscillations
in the CMB spectrum are sensitive to the free-streaming of the neutrinos prior to 
recombination time. The WMAP 7-year data analysis, assuming a $\Lambda$CDM model, 
places a limit $N_\nu >2.7$ (95\% CL) \cite{Larson:2010gs,Komatsu:2010fb}. An earlier analysis 
of the WMAP 5-year data using independent input on the values of the expansion rate 
$H_0$ and the parameter $\sigma_8$ gave a similar limit $N_\nu>2.55$ (95\% CL) 
\cite{Reid:2009nq}. 

Consider the latest limit with $\Delta N_\nu>-0.3$. 
We can replace $T_{\rm nuc}$ in Eq. (\ref{nul})
by the temperature of the universe at matter-radiation equality when the 
density perturbations start to grow, $T_{\rm eq}=0.74{\rm eV}$. The limit on the
tachyon mass becomes
\begin{equation}
\mu\leq 0.33{\rm eV}.
\end{equation}
The limits on the speed of the neutrinos at $28{\rm GeV}$ become
\begin{equation}
v(28{\rm GeV})-1<7.1\times 10^{-23}.
\end{equation}
We also find that $v(10{\rm MeV})-1<5.4\times 10^{-16}$, which is better than
the supernova SN1987A limits on the neutrino velocity, although this only
applies to the case of Lorentz invariant tachyons. With such small tachyonic masses,
the three neutrino species could have a combination of tachyonic and real masses and 
still be consistent with neutrino oscillation experiments 
(which measure differences in $\mu^2$), but this does not affect the bound.

\section{Conclusion}

We conclude that the cosmological bounds rule out the possibility of a tachyonic neutrino
over a wide range of tachyonic masses and the limits are better than those obtained
from supernova 1987A. As a consequence, a GeV energy-scale neutrino that travels appreciably 
faster than the speed of light 
would need to be associated with a violation of Lorentz symmetry. It would be possible to 
take specific models of Lorentz symmetry violation (see \cite{lrr-2005-5,Liberati:2009pf} for reviews)  
and repeat  the cosmological analysis to 
obtain useful limits on these theories.

\bibliography{paper.bib}

\end{document}